\documentstyle[twoside,12pt]{article}
\newcommand{\be}{\begin{equation}}
\newcommand{\ee}{\begin{equation}}
\begin{document}

\begin{center}
{\Large 
{\bf Scaling Exponent for Coarsening in a 1D q-state system.\\ }} \vspace{1in} 
 {\bf  Ajay Gopinathan} \footnote{Present address: Department of Physics, University of Chicago, Chicago, IL 60637, USA}\\
 
\end{center}
  Theoretical Physics Group, Tata Institute of 
Fundamental Research, Colaba, Mumbai, India\\
\vspace{0.5in}
\begin{center}
{\bf Abstract}\\
\end{center}
An exponent $\beta$ which characterises non-equilibrium coarsening 
processes is calculated in a deterministic solvable model of 
coarsening for a 1D q-state Potts system. We study how the 
fraction of sites P which have never changed their state, scale with the characteristic domain length $\langle\ell\rangle$. $\beta$ is defined by  P $\sim \langle \ell \rangle^{\beta -1}$. We 
propose a new model of coarsening that prevents correlations from 
developing between domains thereby ensuring tractability and an exact 
result for any q.\\
\vspace{0.5in}

{\bf1. Introduction\\}

 Domain coarsening  occurs when a system is rapidly quenched from a high 
temperature
 disordered state to a low temperature ordered state [1]. Domains of different 
equilibrium 
states form on larger and larger length scales thus giving us a system 
that is forever in nonequilibrium. An important observation is that this 
nonequilibrium coarsening process exhibits dynamic scaling at late times, 
that is, the domain structure at later times is statistically similar to 
those at earlier times except for a global change of scale [1]. This implies 
that at late times the system is characterized by a {\it single} length 
scale $\langle\ell(t)\rangle$ and all properties of the system can be got in terms of 
this $\langle\ell(t)\rangle$. The scaling hypothesis implies that the equal time and 
two time correlation function can be written in scaling form as
$$
C(r,t) = f(r/\langle\ell\rangle)\hspace{1cm} C(r,t,t') = (\langle\ell\rangle'/\langle\ell\rangle)^{\lambda}h(r/\langle\ell\rangle)\nonumber\\
$$\\
These scaling forms are well supported by experiment.\\

Theoretically there have been two main approaches to studying the 
coarsening process. The first is to study stochastic models such as the 
Ising model with Glauber dynamics and the second is to study 
deterministic models like the Ginzburg-Landau equation for the coarse 
grained order parameter with no noise term. In this case the GL equations 
are completely deterministic, the only randomness being in the initial 
conditions. This is very different from the kinetic Ising models, which 
are stochastic by definition (even at T = 0) and the average of any 
thermodynamic quantity is not only over initial conditions (which is the 
case for the T = 0 GL equations) but also over the histories of 
evolution. It is generally believed that in d$\geq$2 the stochastic and 
deterministic models belong to the same universality class. But in 1D the 
situation is quite different. For example for the 1D Ising 
model, with Glauber dynamics, at T = 0, evolved from a random initial 
condition ( quenching from high T ) , the domain walls behave as independent 
random walkers who annihilate upon meeting. The average domain size 
$\langle\ell\rangle$ grows as t$^{1/2}$. The equal time and the two time 
spin-spin 
correlation functions can be exactly calculated [2,3]. On the other hand for 
deterministic models the situation is different. For example in the 1D GL 
model the domain walls interact via their exponential tails leading to a 
logarithmic growth law [4]. This kind of a difference in time dependence 
suggests that  exponents defined with respect to 
$\langle\ell\rangle$ may be more fundamental than those defined with
respect to $t$.\\

 The correlation function 
exponents and the dependence of the 
characteristic domain length scale on time have been studied extensively [1,2,3,4].
 But since this gives no information on the history of evolution of 
the system, the study of ``persistence" is important. A simple 
way of characterizing the history of evolution of a system would be to 
look at, say, the probability of a single spin in an Ising system not 
changing its spin upto a certain time, in other words, its ``persistence" in 
clinging to its original state. By studying the scaling
behaviour with $\langle\ell\rangle$ of P, the fraction of spins that 
have not changed their state, we can define the exponent $\beta$ by
 P $\sim \langle\ell\rangle^{\beta - 1}$. For the stochastic 1D 
Ising model with 
Glauber dynamics an exact result for $\beta$ was found [5] to be $\beta = 
0.25$. The same exponent was studied in the context of the 1D GL model 
whose late time dynamics can be mapped, as will be explained later, to a 
deterministic coarsening algorithm which consists of the continual 
removal of the smallest domains. In this case it was found [6] to be exactly 
0.824...\\

 The same question can be asked in the context of the q-state Pott's 
model. It turns out that though the autocorrelation exponent and the 
growth law for the characteristic domain length may be the same, $\beta$ 
varies with q. An exact result [5] was recently found for the 1D 
stochastic case, which is an exact generalization of the q = 2 case, as
$$
\beta(q) = \frac{5}{4} - 
\frac{4}{\pi^{2}}\Big[\mbox{cos}^{-1}\Big(\frac{2 - 
q}{\sqrt{2}q}\Big)\Big]^{2}
$$\\

 Noting the difference in value of $\beta$ in the stochastic and 
deterministic case for q = 2, it is only natural to ask how $\beta$ 
varies for general q in the deterministic case.  For general q, however 
tractability is lost because, as we shall explain later, correlations 
develop between domains. Nevertheless approximate calculations 
ignoring these correlations have been carried out [7,8].\\

 In this paper we 
present a different model of coarsening in which no correlations develop 
between domains. Using this model we can calculate $\beta$ exactly for 
any q. For q = 2 our model reduces to the case discussed above.\\

{\bf2. The Model\\ }

The motivation behind our model is as follows.
 Let us consider the q = 2 Ising case first, in 
the context of the time dependent Ginzburg-Landau equation in 1D with no 
noise ( T = 0 ). It turns out [4] that the late time dynamics of this model 
can be mapped to deterministic equations describing the motions of a set 
of interacting domain walls ( kinks in the wall profile ). The kinks 
interact via an exponential attractive interaction. This implies that in 
the limit when the typical separation between domain walls is much larger 
than the width of a wall, the shortest domains collapse instantly as 
compared to the longer ones. This means that the dynamics reduces to a 
deterministic model where at each time step the shortest domain is found 
and the spins inside it are flipped. Thus the system coarsens with the 
continual ``removal" of the shortest domains.\\

  Now we can address the question as to what fraction of the spins have 
never flipped. It is clear that once a domain has been removed the spins 
inside no longer contribute to the fraction of spins that have never 
flipped. If we call those parts of the line where the spins have flipped 
atleast once ``wet", and those parts where the spins have never flipped 
``dry", then clearly, when a domain is removed, the portion of the line it 
occupied becomes ``wet". So the question about the scaling of the fraction 
of spins that have never flipped can be rephrased as: How does the density 
of dry regions scale with $\langle\ell\rangle$? We define the exponent 
$\beta$ by 
saying that the dry region density scales as $\langle\ell\rangle^{\beta - 
1}$ or the dry part per domain goes as $\langle\ell\rangle^{\beta}$.\\

 A straightforward generalization of this procedure to the q - state 
Pott's model can be done. The domains form a random sequence constructed 
from the q available states, no two consecutive domains being the same. 
Again the smallest domains will collapse and can be removed if the states 
of the domains on either side are the {\it same}. But if they are 
{\it different} the two walls will coalesce to form a new wall at the 
midpoint. But this routine leads to correlations developing between the 
domains. Instead we consider an abstract model in which the smallest 
domain is merged with the domain to its right if 
the states of the domains on either side are different. This ensures , as 
we shall shortly explain, that no correlations develop between domains. 
This leads to tractability and an exact result for any q.\\

  In this 
paper we examine how the dry region density scales with the 
characteristic domain length scale in the context of this abstract model.\\

 The details of the model are as follow. \\
  At any time we have a sequence of domains constructed from the q 
available states, no two consecutive ones being the same. The smallest 
domains are identified at each time step. If the domains to its left and 
right are the in the {\it same} state then it flips to {\it that} state 
,that is, the 
three domains merge to form a single large domain. Since domains remain 
uncorrelated, as we shall show, the probability of this occurence is 
$\frac{1}{q -1}$. On the other hand with a probability $\frac{q - 2}{q - 
1}$ ( in the event that the two domains are in different states ), the 
smallest domain flips to the state of its right side neighbour
. In other words, it merges with its right hand
neighbour leaving the other neighbour unaffected. In both cases the part 
of the line previously occupied by the smallest domain becomes wet.\\

 Now we consider the question of correlations. Suppose we choose N 
intervals, the number of distinct arrangements of these on a circle is (N 
- 1)!. Suppose we determistically coarsen this system by picking the 
smallest domain each time and removing it, till we end up with one 
interval. Thus (N - 1)! distinct histories can be created. Now we 
consider an alternate algorithm which consists of picking the smallest 
interval and then picking two intervals at random and combining the 
three. We iterate this procedure until only a single interval remains. 
This procedure also generates (N - 1)! histories which are in one to one 
correspondence with the histories generated by the deterministic 
algorithm described above. This proves that no correlations develop 
between domains during coarsening by merging with domains to {\it both} 
the left and right.\\

 Now consider coarsening the N intervals arranged on a 
circle by picking the smallest interval each time and combining it with 
the interval to the right. This procedure 
generates $(N - 1)!$ histories. As before we consider an 
alternative procedure of picking the smallest interval at each step and 
picking another interval at random and attatching it to the smallest interval.
 This procedure also generates $(N - 1)!$
 histories again proving that the coarsening algorithm does 
not develop any correlations between the domains.\\

 So we have proved that 
the coarsening procedure we use in our model, which is nothing but a 
combination of the two coarsening algorithms described above, does not 
allow correlations to develop between intervals. As a result we can use 
the ``picking at random" algorithm to compute $\beta$.\\

{\bf 3. Equation for $\beta$\\}

 Our calculations follow the method used by Bray {\it et al} [6] for the 1D 
Ising model case. We start with random intervals on a line. Each interval 
{\it I} is characterised by its length {\it l(I)} and by the length of 
its dry part {\it d(I)}. At each time step the smallest domain  
$I_{min}$ is picked.  As explained before there are two possibilities.\\
1. Two more intervals $I_1, I_2$ are picked at random. The three domains are 
merged to form a single large domain $I$. This 
occurs with probability $\frac{1}{q - 1}$. The total length and dry parts 
of {\it I} are given by\\
\begin{equation}
l(I) = l(I_{1}) + l(I_{min}) + l(I_{2})
\end{equation}
\begin{equation}
d(I) = d(I_{1}) + d(I_{2})
\end{equation}\\   
2. Another interval $I_{1}$ is picked at random. The smallest domain is 
merged with it to form a new domain $I$. This occurs with probability 
$\frac{q - 2}{q - 1}$. The length and dry part of $I$ are given by\\
\begin{equation}
l(I) = l(I_{1}) + l(I_{min})
\end{equation}
\begin{equation}
d(I) = d(I_{1})
\end{equation}\\  
 We assume that the lengths of intervals take only integer values and 
that the minimal length in the system is $i_{0}$. We also assume that 
there is a very large number $N$ of intervals. We denote the number of 
intervals of length $i$ by $n_{i}$ and also the average length of 
the dry part of intervals of length $i$ by $d_{i}$. We denote by primed 
symbols the values of these quantities after all the $n_{i_{0}}$ 
intervals of length $i_{0}$ have been eliminated, so that the minimal 
length becomes $i_{0} + 1$. The time evolution equations are then given by\\
\begin{equation}
N' = N - 2n_{i_{0}}(\frac{1}{q-1}) - n_{i_{0}}(\frac{q -2}{q - 1})
\end{equation}\\
i.e\\
\begin{equation}
N' = N - n_{i_{0}}(\frac{q}{q - 1})
\end{equation}\\
Similarly we have\\
\begin{equation}
n'_{i} = n_{i}\Big(1 - \Big(\frac{q}{q - 1}\Big)\frac{n_{i_{0}}}{N}\Big) + 
\frac{n_{i_{0}}}{q - 1}\sum_{j = i_{0}}^{i - 
2i_{0}}\frac{n_{j}}{N}\frac{n_{i - j - i_{0}}}{N} + \frac{q - 2}{q - 
1}n_{i_{0}}\Big(\frac{n_{i - i_{0}}}{N}\Big)
\end{equation}
\begin{eqnarray*}
n'_{i}d'_{i}& = &n_{i}d_{i}\Big(1 - \Big(\frac{q}{q - 
1}\Big)\frac{n_{i_{0}}}{N}\Big) + \frac{n_{i_{0}}}{q - 1}\sum_{j = i_{0}}^{i 
-2i_{0}}\frac{n_{j}}{N}\frac{n_{i - j - i_{0}}}{N}(d_{j} 
+ d_{i - j - i_{0}})\\
\end{eqnarray*}
\begin{equation}
\hspace{5cm} + \Big(\frac{q - 2}{q - 1}\Big)\frac{n_{i_{0}}}{N}(n_{i - 
i_{0}}d_{i - i_{0}})
\end{equation}\\
Note that these are valid for $n_{i_{0}} << N$ which is valid when 
$i_{0}$ becomes large.\\
  We assume that after many iterations {\it i.e.} when $i_{0}$ becomes 
large, a scaling regime is reached, where
\begin{equation}
n_{i} = \frac{N}{i_{0}}f\Big(\frac{i}{i_{0}}\Big)\hspace{2cm}   n_{i}d_{i} = 
Ni_{0}^{\beta - 1}g\Big(\frac{i}{i_{0}}\Big)
\end{equation}\\
 For large $i_{0}$ we can treat $x = i/i_{0}$ as a continuous 
variable. Then neglecting O(1/$i_{0}^2$) we have\\
\begin{equation}
n'_{i} = \frac{N'}{i_{0} + 1}f\Big(\frac{i}{i_{0} + 1}\Big) = 
\frac{N}{i_{0}}\Big[f(x) - \Big(\frac{q}{q - 
1}\Big)\frac{f(1)}{i_{0}}f(x) - \frac{f(x)}{i_{0}} - 
x\frac{f'(x)}{i_{0}}\Big]
\end{equation}\\
and\\
\begin{eqnarray*}
n'_id'_i & = & N(i_{0} + 1)^{\beta - 1}g\Big(\frac{i}{i_{0} + 1}\Big)\\
\end{eqnarray*}
\begin{equation}
\hspace{1cm} = Ni_{0}^{\beta - 1}\Big[g(x) - \Big(\frac{q}{q 
-1}\Big)\frac{f(1)}{i_{0}}g(x) + \frac{\beta - 1}{i_{0}}g(x) 
-x\frac{g'(x)}{i_{0}}\Big]
\end{equation}\\
 Inserting these into the time evolution equations (7) and (8), and using 
the fact that $f$ and $g$ are independent of $i_{0}$ for large $i_{0}$ we 
get
\begin{eqnarray*}
f(x)& +& xf'(x) + \theta(x - 3)\frac{f(1)}{q - 1}\int_{1}^{x - 2}f(y)f(x - 
y - 1)dy\\
\end{eqnarray*}
\begin{equation}
\hspace{1cm} + \theta(x - 2)\Big(\frac{q - 2}{q - 1}\Big)f(1)f(x - 1) = 0
\end{equation}
\begin{eqnarray*}
(1 - \beta)g(x)& +& xg'(x) + 2\theta(x - 3)\frac{f(1)}{q - 1}\int_{1}^{x - 
2}g(y)f(x - y - 1)dy\\
\end{eqnarray*}
\begin{equation}
\hspace{1cm} + \theta(x - 2)\Big(\frac{q - 1}{q - 2}\Big)f(1)g(x 
- 1) = 0
\end{equation}\\
Now we introduce the Laplace tranforms of the functions $f$ and $g$\\
\begin{equation}
\phi(p) = \int_{1}^{\infty}exp(-px)f(x)dx
\end{equation}
\begin{equation}
\psi(p) = \int_{1}^{\infty}exp(-px)g(x)dx
\end{equation}\\
Taking Laplace transforms of equations (12) and (13) we get
\begin{equation}
p\phi'(p) = f(1)exp(-p)\Big[\Big(\frac{1}{q - 1}\Big)\phi^{2}(p) + 
\Big(\frac{q - 2}{q - 1}\Big)\phi(p) - 1\Big]
\end{equation}
\begin{equation}
p\psi'(p) + \beta\psi(p) = f(1)exp(-p)\Big[2\Big(\frac{1}{q - 
1}\Big)\psi(p)\phi(p) + \Big(\frac{q - 2}{q - 1}\Big)\psi(p) - 
\frac{g(1)}{f(1)}\Big]
\end{equation}\\
 These are fairly simple first order differential equations in one 
variable and may be solved in a straightforward fashion to obtain the 
solutions
\begin{equation}
\phi(p) = \frac{1 - exp(-h(p))}{1 + \Big(\frac{1}{q - 1}\Big)exp(-h(p))}
\end{equation}
\begin{equation}
\psi(p) = g(1)\int_{p}^{\infty}\Big[\frac{1 + \frac{1}{q - 
1}exp(-h(x))}{1 + \frac{1}{q - 
1}exp(-h(p))}\Big]^{2}\frac{exp(h(x))}{exp(h(p))}\frac{x^{\beta - 
1}}{p^{\beta}}exp(-x)dx
\end{equation}\\
where h(x) is given by
\begin{equation}
h(x) = f(1)\Big(\frac{q}{q -1}\Big)\int_{x}^{\infty}\frac{e^{-t}}{t}dt
\end{equation}\\
It may be noted that the constants of integration implied by the form of 
the solutions above, are fixed by the requirement that both $\phi$ and 
$\psi$ go to zero as p tends to infinity.\\

It remains to fix the constants f(1) and $\beta$. For this we use the 
following expansion
\begin{equation}
\int_{p}^{\infty}\frac{e^{-x}}{x}dx = -\mbox{ln}(p) - \gamma - \sum_{n = 
1}^{\infty}\frac{(-p)^{n}}{nn!}
\end{equation}\\
where $\gamma$ is Euler's constant and has a value $\gamma = 
0.577215...$. Using this with (18) and (20) gives a small $p$ expansion 
for $\phi$
\begin{equation}
\phi(p) = 1 - \Big(1 + \frac{1}{q - 1}\Big)p^{\frac{q}{q - 
1}f(1)}exp\Big[\frac{q}{q - 1}f(1)\gamma\Big]\Big(1 + O(p)\Big)
\end{equation}\\
Comparing this with the small $p$ expansion from (14) which is $\phi(p) = 
1 - \langle x\rangle p + ...$ we get $f(1) = \frac{q - 1}{q}$ and also 
the ratio of the mean domain length to the minimum length as $\langle 
x\rangle = \Big(\frac{q}{q - 1}\Big)e^{\gamma}$.\\

The exponent $\beta$ can be determined in a similar fashion. We define 
$r(p)$ by 
\begin{equation}
r(p) = h(p) + \mbox{ln}(p) = -\gamma - \sum_{n = 
1}^{\infty}\frac{(-p)^{n}}{nn!}
\end{equation}\\
Using this we can rewrite (19) as\\
\begin{equation}
\psi(p) = g(1)\int_{p}^{\infty}\Big[\frac{1 + \frac{1}{q - 
1}exp(-r(x))}{1 + \frac{1}{q - 
1}exp(-r(p))}\Big]^{2}\frac{exp(r(x))}{exp(r(p))}\frac{x^{\beta - 
2}}{p^{\beta - 1}}exp(-x)dx
\end{equation}\\
From this we may get the small $p$ form of $\psi(p)$ as
\begin{equation}
\frac{g(1)}{1 - \beta} + \frac{g(1)}{1 - \beta}e^{\gamma}p^{1 - 
\beta}B(p,\beta)
\end{equation}\\
where
\begin{equation}
B(p,\beta) = \int_{p}^{\infty}x^{\beta - 
1}\frac{\mbox{d}}{\mbox{dx}}\Big[e^{-x}\Big(e^{\frac{r(x)}{2}} + 
\frac{1}{q - 1}xe^{\frac{-r(x)}{2}}\Big)^{2}\Big]dx
\end{equation}\\
It may be easily shown that
\begin{equation}
B(p,\beta) = B(0,\beta) + O(p^{1 + \beta}) + O(p^{2 + \beta}) + ...
\end{equation}\\
Thus (25) reduces to 
\begin{equation}
\frac{g(1)}{1 - \beta}\Big(1 + B(0,\beta)p^{1 - \beta} + O(p) + O(p^{2}) 
+ ...\Big)
\end{equation}\\
We now compare this expansion to the direct expansion of $\psi(p)$ 
obtained from (15), namely $\psi(p) = \int_{1}^{\infty}dxg(x)(1 - p(x) + 
O(p^{2}))$. If we require the function $g(x)$ to have a finite first 
moment then we have to have $B(0,\beta) = 0$ i.e 
\begin{equation}
 \int_{0}^{\infty}x^{\beta - 
1}\frac{\mbox{d}}{\mbox{dx}}\Big[e^{-x}\Big(e^{\frac{r(x)}{2}} + 
\frac{1}{q - 1}xe^{\frac{-r(x)}{2}}\Big)^{2}\Big]dx = 0
\end{equation}\\
 This condition determines 
$\beta$ for us. For general q the $\beta$ can be determined by numerical 
methods from the above condition (29). We have computed $\beta$ for sample
values of q (see table). 
 $\beta$ for q = 2 agrees with Bray's value [6] upto the 5 decimal
 places we have computed. This is what one would expect because, as we 
 mentioned before, our model reduces to Bray's model for q = 2. We also
 notice a monotonic decrease in $\beta$ as q increases. This is also what one
 would physically expect if we look at the dry part {\it per domain} which scales
 as $\langle\ell\rangle^{\beta}$. As we go to higher values of q, coarsening proceeds
 almost exclusively by the smallest domain merging with {\it one} nearest neighbour.
 This implies that, in the scaling regime, the dry part per domain remains almost 
{\it constant} leading to lower and lower values of $\beta$, which indeed,
 is what we find. It can also be shown analytically by considering the small
 p form of $\psi(p)$ for the q$\rightarrow\infty$ case that $\beta = 0$. \\

\begin{table}[p]
\caption{$\beta$ for various q}
\begin{center}
\begin{tabular}{|c|c|}\hline 
\mbox{\hspace{1cm}q\hspace{1cm}} & \mbox{\hspace{3cm} 
$\beta$\hspace{3cm}} \\ \hline 
2 & $0.82492...$ \\
3 & $0.68092...$ \\
4 & $0.58178...$ \\
5 & $0.50940...$ \\
7 & $0.40999...$\\
10 & $0.31905...$ \\
50 & $0.08321...$ \\ \hline
\end{tabular}
\end{center}
\end{table}

 {\bf4. Conclusion\\}

We have studied a deterministic model of coarsening for the zero temperature
 dynamics of a q-state Pott's system 
in 1D following a rapid quench.
 This model does not allow correlations to develop between domains thus rendering it exactly solvable.
 We have determined the persistence exponent $\beta$ exactly for all q.
  For q = 2 we obtain  $\beta=0.824...$ in complete agreement with [6]. As q increases we find that $\beta$ decays monotonically to zero.\\

\newpage

{\bf Acknowledgements}\\

 I am very grateful to Dr. Satya N. Majumdar who started me off on the problem,
 for his guidance throughout the course of this work. 
 I would also like to thank T.I.F.R, Mumbai for the hospitality extended 
 to me during my stay as a participant in the 
V.S.R.P 97.\\
\newpage
{\bf References}\\

[1] A.J Bray 1994 {\it Adv. Phys.} {\bf 43} 357

[2] R.J. Glauber 1963 {\it J. Mat. Phys.} {\bf 4} 294

[3] A.J. Bray 1990 {\it J. Phys. A} {\bf 23} L67

[4] S.N. Majumdar, D.A. Huse 1995 {\it Phys. Rev. E} {\bf V52 1} 270

[5] B. Derrida, V. Hakim, V. Pasquier 1996 {\it J. Stat. Phys.} {\bf 85} 763 

[6] A.J. Bray, B. Derrida, C. Godreche 1994 {\it Europhys. Lett} {\bf 27} (3) 175

[7] B. Derrida, C. Godreche, I. Yekutieli 1991 {\it Phys. Rev. A} {\bf 44} 6241

[8] P.L. Krapivsky, E. Ben-Naim 1997 {\it Phys. Rev. E} {\bf 56} 3788

\end{document}